\documentstyle[prl,floats,aps,twocolumn,epsf,graphicx]{revtex}
\begin{document}
\twocolumn[\hsize\textwidth\columnwidth\hsize\csname
@twocolumnfalse\endcsname

\title{Time-Dependent Random Walks and the Theory of Complex Adaptive Systems}
\author{Shahar Hod}
\address{Department of Condensed Matter Physics, Weizmann Institute, Rehovot 76100, Israel}
\date{\today}
\maketitle

\begin{abstract}

\ \ \ Motivated by novel results 
in the theory of complex adaptive systems, we analyze the dynamics of random walks 
in which the jumping probabilities are {\it time-dependent}. We determine the 
survival probability in the presence of an absorbing boundary. For an unbiased walk the 
survival probability is maximized in the case of large temporal oscillations in the 
jumping probabilities. On the other hand, a random walker 
who is drifted towards the absorbing boundary performs best with a constant jumping probability. 
We use the results to reveal the underlying dynamics responsible for the phenomenon of self-segregation and 
clustering observed in the evolutionary minority game.
\end{abstract}
\bigskip

]

Random walk is one of the most ubiquitous concepts of statistical physics. In fact, it finds 
applications in virtually every area of physics (see e.g., \cite{BaNi,Kam,FeFrSo,Wei,AvHa,DiDa} and 
references therein). 
Random walks in the presence of absorbing traps are much studied in recent years as models 
for absorbing-state phase transitions \cite{MaDi,Hin}, polymer adsorption \cite{BeLo}, 
granular segregation \cite{FaFu}, and the spreading of an epidemic \cite{GrChRo}.

In this Letter we analyze the problem of a random walk with an absorbing boundary, 
in which the jumping probabilities are 
time-dependent. In addition to the intrinsic interest of such {\it time-dependent} random 
walks, our study is motivated by the flurry of activity in the field of complex adaptive systems.

Agent-based models of complex adaptive systems are attracting significant interest 
across many disciplines \cite{www}. They find applications in social, biological, physical, and economic sciences. 
Of particular interest are situations in which members 
compete for a limited resource, or to be in a minority \cite{John1}. 
In financial markets for instance, more
buyers than sellers implies higher prices, and it is therefore better for a
trader to be in a minority group of sellers. Predators foraging for food will
do better if they hunt in areas with fewer competitors. Rush-hour drivers, facing
the choice between two alternative routes, wish to choose the route containing
the minority of traffic \cite{HubLuk}. 

One of the most studied models in the field of complex adaptive systems is the minority game (MG) 
\cite{ChaZha}, and its evolutionary version (EMG) \cite{John1} 
(see also \cite{DhRo,BurCev,LoHuJo,HuLoJo,HaJeJoHu,BuCePe,LoLiHuJo,LiVaSa,SaMaRi,HodNak1,Coo,HodNak2,CaCe}
and references therein). The game describes agents who each make a binary decision (e.g., ``to buy''/''to sell'', 
or ``taking route A''/``taking route B'') at every point in the game. Profit is made by those who find themselves in the 
minority group, i.e. who end up selling when most wish to buy, or vice versa. Each winner 
gains $R$ points (the ``prize''), while agents belonging to the majority group lose $1$ point (the ``fine'').
The agents have a common ``memory'' look-up table, containing the outcomes of recent occurrences. 
Faced with a given bit string of recent occurrences, 
each agent chooses the outcome in the memory with probability $g$,
known as the agent's ``gene'' value. If an agent score falls below some value $d$, 
then its strategy (i.e., its gene value) is modified. 
In other words, each agent tries to learn from his past mistakes, and
to adjust his strategy in order to survive. 

Early studies of the EMG were restricted to situations in which the prize-to-fine 
ratio $R$ was assumed to be equal unity. 
A remarkable conclusion deduced from the EMG \cite{John1} 
is that a population of competing agents tends to {\it self-segregate} into opposing 
groups characterized by extreme behavior. 
It was realized that in order to flourish in such situations, an agent 
should behave in an extreme way ($g=0$ or $g=1$) \cite {John1}. 

On the other hand, in many real life situations the prize-to-fine 
ratio may take a variety of different values \cite{HodNak1}. A different kind of 
strategy may be more favorite in such situations. 
In fact, we know from real life situations that extreme agents not always perform better than 
cautious ones. 
In particular, our daily experience indicates that in difficult situations 
(e.g., when the prize-to-fine ratio is low) human people tend to be 
confused and indecisive. In such circumstances 
they usually seek to do the same (rather than the opposite) as the majority. 

Based on this qualitative expectation, we have recently extended the 
exploration of the EMG to include situations in which the prize-to-fine ratio $R$ differs from unity. 
It was found \cite{HodNak1} that an intriguing phase transition exist in the model: 
``confusion'' and ``indecisiveness'' take over when the 
prize-to-fine ratio falls below some critical value, 
in which case ``cautious'' agents (characterized by $g={1 \over 2}$) 
perform better than extreme ones. In such circumstances agents tend to 
{\it cluster} around $g={1 \over 2}$ (see Fig. 1 of Ref. \cite{HodNak1}) rather than self-segregate 
into two opposing groups.

Moreover, it has been demonstrated \cite{HodNak1,HodNak2} that the 
population of evolving agents never establishes a genuine stationary 
distribution. In fact, the probability of a particular agent to win oscillates in time. 
This fact has been overlooked in former studies of the EMG. 
The score of each agent may therefore be described by a random walk with {\it time-dependent} 
jumping (winning) probabilities.

Thus, the problem of random walkers whose jumping probabilities are time-dependent is 
of great interest for the understanding of the dynamics of complex adaptive systems. The aim of the 
present Letter is to analyze this problem, and to provide an analytical 
explanation for the phase-transition observed in such systems.

{\it An unbiased random walk.} 
We study a discrete-time random walk on the nonnegative integers, $x_t=0,1,2,\ldots,$ with 
the origin absorbing. Initially the walker is at $x_0=1$ (We shall later generalize our results 
to include situations in which $x_0>1$.) The probability $p(t)$ 
to step to the right is {\it time-dependent} and 
is given by $p(t)={1 \over 2}-(-1)^{t}A$, where $|A|$ is the amplitude 
of the temporal oscillations in the jumping probabilities (${-{1 \over 2}}<A<{1 \over 2}$).

We denote by $y$ the location of the rightmost site yet visited. 
The probability $P(x,y,t)$ for a random walker to be at a position $x$ at time $t$ 
follows the evolution equation

\begin{eqnarray}\label{Eq1}
P(x,y,t+2) =[({1 \over 2}+A)^2+({1 \over 2}-A)^2]P(x,y,t)\nonumber \\  
 +({1 \over 2}+A)({1 \over 2}-A)[P(x-2,y,t)+P(x+2,y,t)] \  ,
\end{eqnarray}
for $x \leq y-2$. We also have $P(0,y,t)=0$, representing the absorbing boundary at the origin. 
Denoting $D(y,t)\equiv P(y,y,t)$, the additional boundary conditions are

\begin{eqnarray}\label{Eq2}
D(y,t+2)& = & ({1 \over 4}-A^2)[P(y-2,y,t)+P(y-2,y-1,t)\nonumber \\
 &&+D(y-2,t)]+[{1 \over 2}+(-1)^yA]^2 D(y,t)\  ,
\end{eqnarray}

\begin{eqnarray}\label{Eq3}
P(y-2,y,t+2)& = & ({1 \over 4}-A^2)[P(y-4,y,t)+D(y,t)]\nonumber \\
 &&+2({1 \over 4}+A^2)P(y-2,y,t)\  ,
\end{eqnarray}
and

\begin{eqnarray}\label{Eq4}
P(y-2,y-1,t+2)& = & ({1 \over 4}-A^2)P(y-4,y-1,t)\nonumber \\
 &&+2({1 \over 4}+A^2)P(y-2,y-1,t)\nonumber \\
 &&+[{1 \over 2}-(-1)^yA]^2D(y-2,t)\  .
\end{eqnarray}

It proofs useful to define the generating function $\hat P(x,y,z)=\sum ^{\infty }_{t=0} z^t P(x,y,t)$ 
(and similarly for D). Multiplying Eqs. (\ref{Eq1})-(\ref{Eq4}) by $z^t$ 
and summing over $t$, one finds that $\hat P(x,y,z)$ satisfies the recursion relation

\begin{eqnarray}\label{Eq5}
z^{-2} \hat P(x,y)& = & ({1 \over 4}-A^2)[\hat P(x-2,y)+\hat P(x+2,y)]\nonumber \\
 &&+2({1 \over 4}+A^2) \hat P(x,y)\  ,
\end{eqnarray}
(we drop the argument $z$ for brevity), subject to the boundary conditions 

\begin{eqnarray}\label{Eq6}
z^{-2} \hat D(y)& = & ({1 \over 4}-A^2)[\hat P(y-2,y)+\hat P(y-2,y-1)\nonumber \\
 &&+\hat D(y-2)]+[{1 \over 2}+(-1)^yA]^2 \hat D(y)\  ,
\end{eqnarray}

\begin{eqnarray}\label{Eq7}
z^{-2} \hat P(y-2,y)& = & ({1 \over 4}-A^2)[\hat P(y-4,y)+\hat D(y)]\nonumber \\
 &&+2({1 \over 4}+A^2)\hat P(y-2,y)\  ,
\end{eqnarray}
and

\begin{eqnarray}\label{Eq8}
z^{-2} \hat P(y-2,y-1)& = & ({1 \over 4}-A^2) \hat P(y-4,y-1)\nonumber \\
 && +2({1 \over 4}+A^2) \hat P(y-2,y-1)\nonumber \\
 && +[{1 \over 2}-(-1)^yA]^2 \hat D(y-2)\  .
\end{eqnarray}

The solution of Eq. (\ref{Eq5}) subject to the boundary condition at the origin, $\hat P(0,y)=0$, 
is $\hat P(x,y)=(\lambda^x- \lambda^{-x}) \hat B(y)$, where $\lambda$ is a function of 
$z$ and $A$. To determine the survival probability as $t \to \infty$, we shall analyze the behavior 
of $\hat P(x,y,z)$ as $z \to 1$. In this limit 

\begin{equation}\label{Eq9}
\hat C(x) \equiv \lambda^x- \lambda^{-x}=2 \sinh \left[\sqrt{{\epsilon} \over {2({1 \over 4}-A^2)}}x \right]\  ,
\end{equation}
where $\epsilon=1-z$.

Substituting Eqs. (\ref{Eq7}) and (\ref{Eq8}) in Eq. (\ref{Eq6}) one obtains the recursion relation 

\begin{eqnarray}\label{Eq10}
\left[(2+q) {\hat C(y-2) \over \hat C(y-4)} -(1+q) \right] \hat B(y)={1 \over 2}q \hat B(y-1) \nonumber \\ 
 +{1 \over 2}(1+q){\hat C(y-2) \over \hat C(y-4)} \hat B(y-2) \  ,
\end{eqnarray}
whose asymptotic solution is 

\begin{equation}\label{Eq11}
\hat B(y) =  {{\sqrt{s \epsilon}} \over {\sinh^2 (\sqrt{s \epsilon}y)}}
{{{1 \over 2}+A} \over {{1 \over 2}+(-1)^yA}}\  ,
\end{equation}
where $q \equiv {1 \over 2}+(-1)^yA$ and $s \equiv {1 \over {2({1 \over 4}-A^2)}}$.

The survival probability is determined by 
$S(t)=\sum ^{\infty }_{y=0} \sum ^{y}_{x=0}P(x,y,t)$. Taking cognizance of 
Eqs. (\ref{Eq9}) and (\ref{Eq11}) one finds that the singular behavior of its 
generating function, $\hat S(z)$, is dominated by 

\begin{equation}\label{Eq12}
\hat S(z)=\sum ^{\infty }_{y=0} \hat B(y) \sum ^{y}_{x=0} \hat C(x)= 
\sqrt{2({1 \over 2}+A) \over {{1 \over 2}-A}} \epsilon^{-{1 \over 2}} \  .
\end{equation}
Nothing that for large $t$, the coefficient of $z^t$ in $(1-z)^{-{1 \over 2}}$ is 
$1 /{\sqrt{\pi t}}$, we have that the asymptotic behavior of the survival probability is 

\begin{equation}\label{Eq13}
S(t;x_0=1)=\sqrt{{1 \over 2}+A \over {{1 \over 2}-A}} \sqrt{2 \over \pi} t^{-{1 \over 2}}\  .
\end{equation}

Finally, it is straightforward to prove the recursion relations 
$S(t;x_0=2n+1,A)=(1+ {n \over {{1 \over 2}+A}}) S(t;x_0=1,A)$ and 
$S(t;x_0=2n,A)=({n \over {{1 \over 2}-A}}) S(t;x_0=1,-A)$. Thus, the 
survival probability for an arbitrary value of the initial position $x_0$ is given by 

\begin{equation}\label{Eq14}
S(t)={{x_0+[1-(-1)^{x_0}]A} \over \sqrt{1-4A^2}} 
\sqrt{2 \over \pi} t^{-{1 \over 2}}\  .
\end{equation}
We therefore conclude that an unbiased random walker whose jumping probabilities display 
{\it large} temporal oscillations 
(i.e., $|A| \simeq {1 \over 2}$) has the largest survival probability. 
On the other hand, the survival probability $S(t)$ has a minimum 
at $A=0$ for even values of $x_0$ (corresponding to a fixed, time-independent jumping probability), 
and at $A={{-1} \over {2x_0}}$ for odd values of $x_0$.

Figure 1 displays the survival probability 
as a function of $A$, the amplitude of the temporal oscillations in the jumping probabilities. 
We find an excellent agreement between the analytically predicted results [see Eq. (\ref{Eq14})] 
and the numerical ones.

\begin{figure}[tbh]
\centerline{\epsfxsize=9cm \epsfbox{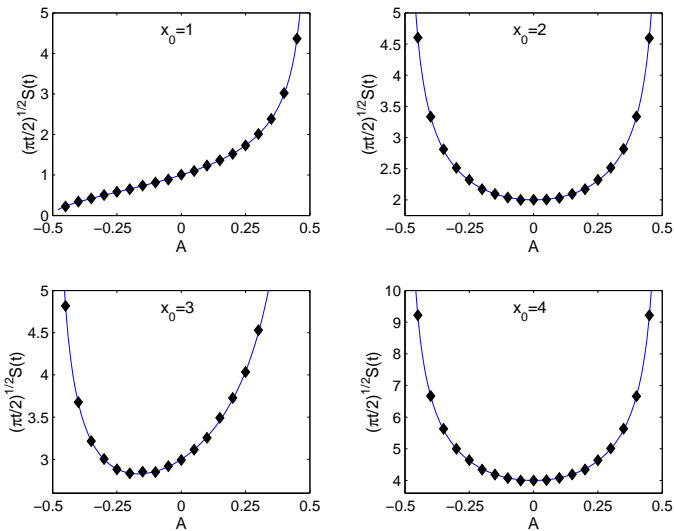}} 
\caption{The survival probability $S(t;x_0) {\sqrt{\pi \over 2}}t^{1 \over 2}$ as a function of 
the amplitude of the temporal oscillations in the jumping probabilities. We display results for 
different values of the initial location: $x_0=1, x_0=2, x_0=3$, and $x_0=4$. Theoretical results 
are represented by solid curves.}
\label{Fig1}
\end{figure}

{\it A biased random walk.}
We next consider situations in which the step length to the left 
is larger than the step length $R$ to the right (where $R \leq 1$). 
Thus, the walker is drifted towards the 
absorbing boundary. We shall also generalize the analysis to 
include situations in which the jumping probability to the left is different from the 
corresponding jumping probability to the right. 
The ({\it time-dependent}) jumping probability to the right is now given by 
$p(t)={1 \over 2}-\varepsilon-(-1)^t A$, where ${-{1 \over 2}}+\varepsilon<A<{1 \over 2}-\varepsilon$. 

In order to survive under such conditions, 
the walker must step to the right more times than he steps to the left. More precisely, at 
least a fraction $f={1 \over {1+R}}$ of his steps must be to the right (at any given point in 
the walk). The chance for the mean number of right-steps to deviate from its average is 
exponentially decreasing with time. Using an analysis along the same lines as before, one finds that the 
asymptotic large $t$ limit of the survival probability is now given by \cite{Note1}

\begin{equation}\label{Eq15}
S(t)=bt^{-{3 \over 2}} e^{-{Ft}}\  ,
\end{equation}
where the associated entropy function (or large-deviation function) reads

\begin{eqnarray}\label{Eq16}
F={1 \over 2} \Big[f_+ \ln\left({f_+ \over p_+}\right) +(1-f_+) 
\ln\left({{1-f_+} \over {1-p_+}}\right)\nonumber \\ 
+f_- \ln\left({f_- \over p_-}\right) +(1-f_-) 
\ln\left({{1-f_-} \over {1-p_-}}\right)\Big]\  ,
\end{eqnarray}
with $p_{\pm}={1 \over 2}-\varepsilon \pm A$, and 

\begin{eqnarray}\label{Eq17}
f_{\pm} & = & f \pm \Big[1-4\varepsilon^2+4A^2 \nonumber \\
&& -\sqrt{(1-4\varepsilon^2+4A^2)^2-64f(1-f)A^2}\Big]/8A\  .
\end{eqnarray}
[The explicit expression of the prefactor $b=b(R)$ is not important for the analysis].

The survival probability at late times is dominated by the exponential factor 
$ e^{-{Ft}}$. 
We note that $e^{-F}$ [and thus also $S(t)$] is a monotonic decreasing function 
of $|A|$. One therefore concludes that a {\it biased} random walker (one 
who is drifted towards the absorbing boundary) with a constant 
jumping probability (i.e., with $A=0$) 
has the largest survival probability. On the other hand, 
a (biased) random walker with large temporal oscillations in his jumping probabilities 
has the poorest chances of survival.

{\it Complex adaptive systems.} 
We now apply the results derived in the present work to provide an explanation for the intriguing 
phase-transition observed in the EMG [transition from self-segregation to clustering (with an 
intermediate M-shaped phase), as the value of the 
prize-to-fine ratio $R$ falls below some critical value].

As mentioned, it has been demonstrated \cite{HodNak1} that the 
population of evolving agents never establishes a true stationary 
distribution. In fact, the probability of a particular agent to win (to step to the right in the 
terminology of this Letter) {\it oscillates} in time, the amplitude being 
dependent on the particular gene-value of the agent \cite{HodNak2}. 
``Extreme'' agents (with $g=0,1$) have an oscillation amplitude which is larger than the 
corresponding amplitude of ``cautious'' agents (those with $g={1 \over 2}$). 
In fact, the winning probability of $g \simeq {1 \over 2}$ agents is 
practically constant in time. We therefore write $A^2=a^2(R) (g-{1 \over 2})^{2}$ \cite{Note2}.

One should also take into account the self-interaction (or so-called {\it market impact} in financial market 
terminology) that agents in such systems experience \cite{LoHuJo}. An agent has a {\it smaller} 
probability of winning when participating in the game as compared to an outsider, someone whose action 
does not affect the outcome. The self-interaction term has the form 
$\varepsilon(g)={\varepsilon_0 \over \sqrt{N}}g(1-g)$ \cite{LoHuJo}.

Substituting the expressions for $A(g)$ and $\varepsilon(g)$ 
in Eq. (\ref{Eq16}) one finds that $e^{-F}$ 
[and thus also $S(t)$] has three qualitatively 
different forms, depending on the precise value of $R$ 
(the step-lengths ratio, or equivalently the prize-to-fine ratio). 
Figure 2 displays the function $e^{-F}$, which determines the long-time survival probability of 
the agents. This figure demonstrates the 
phase-transition from self-segregation to clustering observed in the 
evolutionary minority game \cite{John1,HodNak1} (compare, in particular, with the numerical results 
presented in Fig. 1 of \cite{HodNak1}).

\begin{figure}[tbh]
\centerline{\epsfxsize=9cm \epsfbox{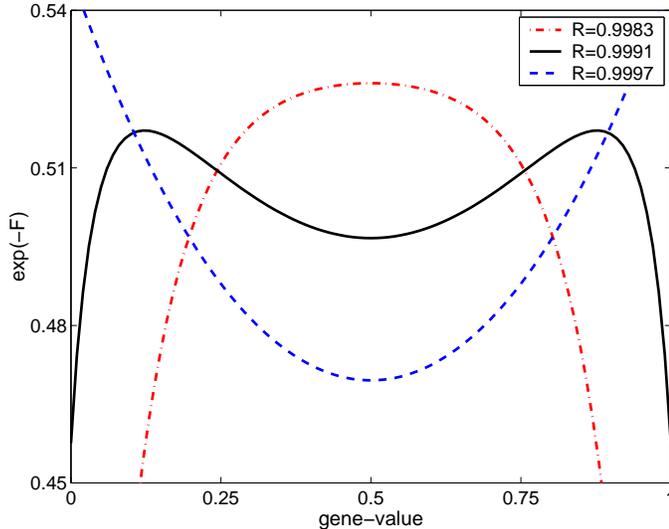}} 
\caption{The survival probability for different values of the 
prize-to-fine ratio: $R=0.9983, R=0.9991$, and $R=0.9997$. The parameters 
used in the figure are ${{\varepsilon_0}/{\sqrt{N}}}=10^{-3}$ and $a=0.9$. Compare this figure 
with Fig. 1 of [25]. The graphs are rescaled for convenience.}
\label{Fig2}
\end{figure}

The phase-transitions are described by the following critical values:

\begin{equation}\label{Eq18}
R^{(1)}_c=1-({2 \over a^2}-2){\varepsilon_0 \over \sqrt{N}}\ \ \ ; \ \ \ 
R^{(2)}_c=1-({2 \over a^2}-1){\varepsilon_0 \over \sqrt{N}}\  .
\end{equation} 
For $R>R^{(1)}_c$ the survival probability $S(t)$ is peaked around $g=0$ and 
$g=1$, and has a minimum at $g={1 \over 2}$. To survive under such conditions, 
an agent (a biased random walker) should 
behave in an {\it extreme} way (that is, should have large temporal oscillations in his 
jumping probabilities). On the other hand, for $R<R^{(2)}_c$ [poor conditions, in which 
the fine (length step to the left) is larger than the reward (length step to the right)] 
one finds that $S(t)$ is peaked around $g={1 \over 2}$. This corresponds to {\it cautious} agents 
with constant (time-independent) jumping probabilities. 
Thus, cautious agents out-perform the extreme ones under harsh conditions. 
[It should be emphasized that this occurs despite the fact that the average winning probability of 
extreme agents (${1 \over 2}$) is actually larger than the corresponding 
probability of cautious agents  (${1 \over 2} -\varepsilon$)].
 
There is also an intermediate phase [for $R^{(2)}_c < R < R^{(1)}_c$], in which the 
survival probability has an M-shaped distribution. This behavior is a direct 
consequence of two {\it opposing} factors: (a) the fact that a biased random walker 
with a constant jumping probability ($A=0$, or equivalently $g={1 \over 2}$) 
has the largest survival probability, and 
(b) the fact that the market impact (which decreases the winning probability) is the largest 
for $g={1 \over 2}$ agents. 
Note that the phase-transition from self-segregation to clustering becomes sharp as the 
number of agents $N$ increases. In fact, the intermediate (M-shaped) phase disappears in 
the $N \to \infty$ limit, in which case the transition occurs at $R^{(1)}_c =R^{(2)}_c=1$ [see 
Eq. (\ref{Eq18})].

In summary, in this Letter we have analyzed the dynamics of random walks with time-dependent 
jumping probabilities. In particular, we have calculated the survival probability 
of such walkers in the presence of an absorbing boundary. It was shown that the best strategy to be adopted by the 
walkers depends on the precise value of the step-lengths ratio $R$. 
In the unbiased case ($R=1$) the survival probability is maximized by a walker 
who has {\it large} temporal oscillations in his jumping probabilities. On the other hand, a random 
walker who is drifted towards the absorbing boundary is better off keeping his jumping probabilities constant 
(i.e., time-independent).

Furthermore, we have shown that the results, when applied to the theory of 
complex adaptive systems, provide a direct analytical explanation for the phase-transition 
(from {\it self-segregation} to {\it clustering}) observed in the evolutionary 
minority game.
        
\bigskip
\noindent
{\bf ACKNOWLEDGMENTS}
\bigskip

I thank Mordehai Milgrom for his kind assistance, and a support by the 
Dr. Robert G. Picard fund in physics. I would also like to 
thank Ehud Nakar, Oded Hod, Assaf Pe`er and Clovis Hopman for helpful discussions. 
This research was supported by grant 159/99-3 from the Israel Science Foundation.


\begin{thebibliography}{99}

\bibitem{BaNi} M. N. Barber and B. W. Ninham, {\it Random and Restricted Walks} (Gordon 
and Breach, New York, 1970).

\bibitem{Kam} N. G. van Kampen, {\it Stochastic Processes in Physics and 
Chemistry} (North-Holland, Amsterdam, 1992).

\bibitem{FeFrSo} R. Fernandez, J. Frohlich, and A. D. Sokal, {\it Random Walks, Critical 
Phenomena, and Triviality in Quantum Field Theory} (Springer Verlag, Berlin, 1992).

\bibitem{Wei} G. H. Weiss, {\it  Aspects and Applications of the Random Walk} (North 
Holland, Amsterdam, 1994).

\bibitem{AvHa} D. ben-Avraham and S. Havlin, {\it Diffusion and Reactions in Fractals and 
Disordered Systems} (Cambridge University Press, Cambridge, 2000).

\bibitem{DiDa} R. Dickman and D. ben-Avraham, Phys. Rev. E. {\bf 64}, 020102(R) (2001).

\bibitem{MaDi} J. Marro and R. Dickman, {\it Nonequilibrium Phase Transitions in Lattice Models}, 
(Cambridge University Press, Cambridge, 1999).

\bibitem{Hin} H. Hinrichsen, Adv. Phys. {\bf 49}, 815 (2000).

\bibitem{BeLo} K. De'Bell and T. Lookman, Rev. Mod. Phys. {\bf 65}, 87 (1993).

\bibitem{FaFu} Z. Farkasa and T. Fulop, J. Phys. A: Math. Gen. {\bf 34}, 3191 (2001).

\bibitem{GrChRo} P. Grassberger, H. Chate, and G. Rousseau, Phys. Rev. E {\bf 55}, 2488 (1997).


\bibitem{www} For a detailed account of previous work on agent-based models such as the Minority Game 
see http://www/unifr.ch/econophysics.

\bibitem{John1} N. F. Johnson, P. M. Hui, R. Jonson, and T. S. Lo, Phys. Rev. 
Lett. {\bf 82}, 3360 (1999).

\bibitem{HubLuk} B. Huberman and R. Lukose, Science {\bf 277}, 535 (1997).

\bibitem{ChaZha} D. Challet and C. Zhang, Physica A {\bf 246}, 407 (1997); 
{\bf 256}, 514 (1998); {\bf 269}, 30 (1999).

\bibitem{DhRo} R. D`Hulst and G. J. Rodgers, Physica A {\bf 270}, 514 (1999).

\bibitem{BurCev} E. Burgos and H Ceva, Physica {\bf 284A}, 489 (2000). 

\bibitem{LoHuJo} T. S. Lo, P. M. Hui and N. F. Johnson, Phys. Rev. E {\bf 62}, 4393 (2000).

\bibitem{HuLoJo}  P. M. Hui, T. S. Lo, and N. F. Johnson, e-print cond-mat/0003309.

\bibitem{HaJeJoHu} M. Hart, P. Jefferies, N. F. Johnson and P. M. Hui, e-print cond-mat/0003486; 
e-print cond-mat/0004063.

\bibitem{BuCePe} E. Burgos, H. Ceva and R. P. J. Perazzo, e-print cond-mat/0007010.

\bibitem{LoLiHuJo} T. S. Lo, S. W. Lim, P. M. Hui and N. F. Johnson, Physica {\bf 287A}, 313 (2000).

\bibitem{LiVaSa} Y. Li, A. VanDeemen and R. Savit, e-print nlin.AO/0002004.

\bibitem{SaMaRi} R. Savit, R. Manuca and R. Riolo, Phys. Rev. Lett. {\bf 82}, 2203 (1999).

\bibitem{HodNak1} S. Hod and E. Nakar, Phys. Rev. Lett. {\bf 88}, 238702 (2002).

\bibitem{Coo} A. C. C. Coolen, e-print cond-mat/0205262.

\bibitem{HodNak2} E. Nakar and S. Hod e-print cond-mat/0206056; Phys. Rev. E (to be published).

\bibitem{CaCe} I. Caridi and H. Ceva, e-print cond-mat/0206515.

\bibitem{Note1} This result can also be obtained using the Sparre Andersen combinatorial relation 
\cite{Fel,Spa}. See \cite{BaGoLu} for more details in the simpler case of time-independent 
jumping probabilities.

\bibitem{Fel} W. Feller, {\it An Introduction to Probability Theory and its Applications}, 
Volumes 1 \& 2 (Wiley, New York, 1968, 1971).

\bibitem{Spa} E. Sparre Andersen, Math. Scand. {\bf 1}, 263 (1953); {\bf 2}, 195 (1954).

\bibitem{BaGoLu} M. Bauer, C. Godreche and  J. M. Luck, J. Stat. Phys. {\bf 96}, 963 (1999).

\bibitem{Note2} One can use a more general expansion of the form 
$A^2=\sum ^{\infty}_{k=1} a_k (g-{1 \over 2})^{2k}$, but 
this would not change the final conclusions to be discussed below.

\end{thebibliography}
\end{document}